\documentclass[12pt]{article}
\usepackage{epsfig}

\parskip=\medskipamount	% space between paragraphs (TeX)
\renewcommand{\thesection}{\arabic{section}}

\catcode`@=11

\@addtoreset{equation}{section}
\@addtoreset{equation}{subsection}
\def\theequation{\ifnum\value{section}=0 \arabic{equation}\ignorespaces
\else \ifnum\value{section}=-1 A.\arabic{equation}\ignorespaces
\else \ifnum\value{subsection}=0 \thesection.\arabic{equation}\ignorespaces
\else \thesection.\arabic{subsection}.\arabic{equation}\ignorespaces
                             \fi
                        \fi
                   \fi}

{\catcode`\'=\active \def'{{}^\bgroup\prim@s}}

\catcode`@=12

\newcommand{\bq}{\begin{equation}}
\newcommand{\be}{\begin{equation}} 
\newcommand{\fq}{\end{equation}}
\newcommand{\ee}{\end{equation}}
\newcommand{\bqr}{\begin{eqnarray}}
\newcommand{\beqs}{\begin{eqnarray}} 
\newcommand{\fqr}{\end{eqnarray}}
\newcommand{\eeqs}{\end{eqnarray}}

\newcommand{\rf}[1]{(\ref{#1})}

 \def\cN{{\cal N}}

\def\pa{\partial}

\def\on#1#2{{\buildrel{\mkern2.5mu#1\mkern-2.5mu}\over{#2}}}
\def\dt#1{\on{\hbox{\bf .}}{#1}}                % (big) dot over

\def\a{\alpha} 
\def\b{\beta} 
\def\da{\dt\alpha}
\def\db{\dt\beta}

\def\bop#1{\setbox0=\hbox{$#1M$}\mkern1.5mu
	\vbox{\hrule height0pt depth.04\ht0
	\hbox{\vrule width.04\ht0 height.9\ht0 \kern.9\ht0
	\vrule width.04\ht0}\hrule height.04\ht0}\mkern1.5mu}
\def\Box{{\mathpalette\bop{}}}                        % box

\begin{document}
\thispagestyle{empty} 

\begin{flushright} 
\begin{tabular}{l} 
ANL-HEP-PR-01-016 \\
hep-th/0103132 \\ 
\end{tabular} 
\end{flushright}  

\vskip .3in 
\begin{center} 

{\Large\bf Massive selfdual perturbed gauge theory} 

\vskip .3in 

{\bf Gordon Chalmers} 
\\[5mm] 
{\em Argonne National Laboratory \\ 
High Energy Physics Division \\ 
9700 South Cass Avenue \\ 
Argonne, IL  60439-4815 } \\  

{e-mail: chalmers@pcl9.hep.anl.gov}  

\vskip .5in minus .2in 

{\bf Abstract}  

\end{center}
\noindent
Spontaneously broken gauge theories are described as a perturbation of selfdual 
gauge theory.  Instead of the incorporation of scalar degrees of freedom, the 
massive component of the gauge field is obtained from an anti-selfdual field 
strength consisting of three components before gauge fixing.  The interactions 
describe a massive gauge theory that is non-polynomial with an expansion containing 
an infinite number of terms.  The Lagrangian generalizes the form of the axial anomaly 
in two dimensions.  Unitary propagation of the tensor field occurs 
upon gauge fixing an additional symmetry.

\setcounter{page}{0} 
\newpage 
\setcounter{footnote}{0} 

\section{Introduction} 

Selfdual quantum field and string theories describe a sector of both gauge and gravity 
theories \cite{Chalmers:1996rq}.  Perturbations of these selfdual models generate 
reformulations of quantum gauge theories that have simplified expansions about the 
selfdual limit \cite{Chalmers:1999jb,Morgan:1995te,Chalmers:1997sg,Chalmers:2001cy}.  
In this work, we examine a selfdual inspired reformulation of quantum gauge theory 
possessing a covariant mass term that is renormalizable and unitary.  Selfdual systems 
are quasi two-dimensional in the sense that the analytic properties of the scattering 
are contained in one half of the Lorentz group.  The mass term that we examine in this 
work is a generalization of the integrated axial anomaly in two dimensions; it has a 
local description via the implementation of auxiliary fields, describing a perturbation 
of selfduality.  

A Lorentz covariant form of selfdual quantum field theory is described by the lagrangians 
${\cal L} = {\rm Tr} ~G^{\a\b} F_{\a\b}$ for spin one and ${\cal L} = \rho^{\a\b} R_{\a\b}$ for 
spin two theories (proposed in \cite{Siegel:1992xp} and quantized in \cite{Chalmers:1996rq}).  
We adopt the conventions of those in \cite{superspaceconventions}.  Non-lorentz covariant 
versions \cite{sdfield} involving one degree of freedom give identical one-loop results (modulo 
a factor of half) but are inconsistent at higher loop orders.  The selfdual field strengths 
are the selfdual projections with $R_{\a\b}$ the selfdual component of the Weyl tensor.  The 
quantum theories may be solved exactly in perturbation theory with the former described by the 
S-matrix \cite{Bern:1994qk,Mahlon:1994si}, 
\bqr  
A_{n;1}^{[J]} = -{i\over 48\pi^2} 
\sum_{1\leq i<j<k<l\leq n} {\langle ij\rangle [jk]\langle kl\rangle 
 [ln]\over \langle 12\rangle \langle 23\rangle \cdot\cdot\cdot \langle n1\rangle } \ , 
\label{MHV} 
\fqr 
with $[J]$ denoting the spin of the internal virtual particle\footnote{Reviews of the 
techniques in calculating gauge theory amplitudes are found in \cite{Mangano:1991by} for 
tree-level processes and in \cite{Bern:1996je} at the one-loop level.}.  The tree-level 
amplitudes of the selfdual gauge theory are equal to zero except for a nonvanishing 
three-point vertex.   The expression is written in spinor helicity form 
\cite{spinorone,spinortwo,oldtwist}, in which $\langle ij\rangle = k_i^\a k_{j,\b}$ and 
$[ij]=k_i^{\da} k_{j,\da}$ are inner products with respect to the covering of the Lorentz 
group; they are complex conjugates in $d=3+1$ dimensions, but real and independent 
in $d=2+2$ dimensions.  The subscript on $A_{n;1}^{[J]}$ denotes  the leading in color 
component of the quantum amplitude in color-ordered form \cite{Bern:1991ux,Bern:1996je}, 
and is expressed in terms of the basis,  
\bqr  
N^2 {\rm Tr} ~T^{a_1} T^{a_2} \cdot T^{a_n} ,  
\fqr 
for an internal state of spin $[J]$ in the adjoint representation.  Sub-leading in 
color partial amplitudes may be deduced via permutations of the indices of the former 
\cite{Bern:1995fz}.  A similar exact expression follows for the gravitational theory 
\cite{Bern:1998xc}.  Furthermore, these S-matrices have a description in terms of the 
$N=2$ quantum string theory defined by self-consistency between different orders in the 
loop expansion \cite{Chalmers:2000bh} rather than the usual path integral quantization 
\cite{Ooguri:1991fp,Bischoff:1997bn,Chalmers:2000wd,Marcus:1992xt}.  Several exact sequences 
of amplitudes have compact forms, including the tree-level next-to-maximal Parke-Taylor 
helicity violating amplitudes \cite{ptamps} and those in supersymmetric theories 
\cite{Bern:1994zx,Bern:1995cg}.  

The description of quantum gauge theory may be simplified by perturbing the selfdual 
gauge theory by dualizing the theory with $G^{\a\b}$ transforming as the anti-selfdual 
$(0,1)$ tensor \cite{Chalmers:1997sg,Freedman:1981us}, 
\bqr  
{\cal L} = {\rm Tr} \left( -{1\over 2} G^{\a\b} G_{\a\b} + G^{\a\b} F_{\a\b} \right) \ ,
\fqr 
which generates ${\rm Tr} F^2$.  The perturbations of this theory and its massive version 
describing spontaneously broken gauge theories \cite{Chalmers:2001cy} generate simplified 
Feynman diagrammatics when expanded around the selfdual helicity configuration;  This 
expansion is one in helicity flips around the maximally helicity violating amplitude in 
\rf{MHV}.  Interestingly, these S-matrices also have a description in terms of the $\cN=2$ 
quantum string theory (and non-maximal helicity violating amplitudes may be obtained by 
deformations) and maps a direct perturbative QCD/string relation \cite{Chalmers:2000bh}.  
In this work we examine more general deformations of selfdual quantum field theories that 
describe massive gauge theories.  As the Higgs particle has not been found it is of interest 
to generate alternative means to describe massive vector theories.  

This work is organized as follows.  In section 2 we dualize the massive theory beginning 
with a local representation of the theory and generate the interactions, comparing with 
previous work.  In section 3 we list the diagrammatic rules and analyze briefly the 
renormalizability and unitarity properties of the theory.  

\section{Dual Lagrangian} 

Self-dual field theories, quantum field theories with selfdual equations of motion, have been 
extensively studied and multiple actions have been proposed.  A Lorentz covariant 
version is described by the Lagrangian, 
\bqr  
{\cal L} = {\rm Tr}~ G^{\a\b} F_{\a\b} \ , 
\fqr 
and admits an exact solution which consists of a one-loop amplitude together with a 
non-vanishing three-point tree vertex.  Gauge theory in general may be dualized by 
expressing the theory via 
\bqr  
{\cal L} = {1\over 2} {\rm Tr} ~F^{\a\b} F_{\a\b} \rightarrow 
 {\cal L} = {\rm Tr} \Bigl( -{1\over 2} G^{\a\b} G_{\a\b} + G^{\a\b} F_{\a\b} \Bigr) \ , 
\label{deformedgauge}
\fqr 
that is, as a deformation around the self-dual sector.  More general deformations of 
self-duality are also obtained and may be used to formulate a massive version of gauge theory.  

The modified selfdual gauge theory we examine is given by, 
\bqr  
{\cal L} = {\rm Tr}~ {1\over 2} G^{\a\b} \Box G_{\a\b} + M G^{\a\b} F_{\a\b} + 
{1\over 2} G^{\da\db} \Box G_{\da\db} + MG^{\da\db} F_{\da\db} 
\nonumber 
\fqr 
\bqr 
+ {1\over 2} {\rm Tr} ~\left( F^{\a\b} F_{\a\b} + F^{\da\db} F_{\da\db}\right) \ , 
\label{localform}
\fqr 
with the field strengths defined as $F_{\a\b} = i{1\over 2}\partial_{(\a}^\da A_{\b)}{}^\da+ 
A_{(\a}{}^\da A_{\b\da}$ which differs from \ref{deformedgauge} via a covariant 'de 
Lambertian on the dual fields\footnote{The form in four-component vector notation follows from 
replacing $G_{\a\b}\rightarrow G_{\mu\nu}+i{\tilde G}_{\mu\nu}$ and $G_{\da\db}\rightarrow 
G_{\mu\nu}-i{\tilde G}_{\mu\nu}$ with ${\tilde G}_{\mu\nu}={i\over 2} 
\epsilon_{\mu\nu\rho\gamma} 
G^{\rho\gamma}$.}.  As the theory in \rf{localform} has terms of mass dimension 
less than or equal to four the theory is renormalizable.  To implement the dualization 
we need to add in a counterterm $Z_m m^2/2 (G^{\a\b}G_{\a\b}+G^{\da\db}G_{\da\db})$ so 
that the renormalized mass is equal to zero.  The theory in \rf{localform} is distinguished 
from a propagating Kalb-Ramond field with kinetic term ${\cal L} = -1/4 H^{\mu\nu\rho} 
H_{\mu\nu\rho}$ and field strength $H_{\mu\nu\rho} = \partial_\mu G_{\nu\rho} + \partial_\nu 
G_{\mu\rho} + \partial_\rho G_{\mu\nu}$.  

The covariant derivatives are defined as, 
\bqr 
\Box = \pa^{\a\da}\pa_{\a\da} + \left\{ A^{\a\da},\pa_{\a\da} \right\} + A^{\a\da}A_{\a\da} \ , 
\fqr 
which requires expansions in the forms of the interactions that follow in the denominator.  
The Lagrangian has equations of motion, 
\bqr 
\Box G^{\a\b} + MF^{\a\b} = 0 \qquad G^{\a\b} = -{M\over \Box} F_{\a\b} \ , 
\fqr 
which generates the classical theory, 
\bqr 
{\cal L} = {\rm Tr} \Bigl( {1\over 2} F^{\a\b} F_{\a\b} + {1\over 2} F^{\da\db}F_{\da\db} 
 -{M^2\over 2} F^{\a\b}{1\over \Box} F_{\a\b} - {M^2\over 2} F^{\da\db} {1\over\Box} 
F_{\da\db} \Bigr)  \ ,
\label{massivetheory} 
\fqr 
with a covariant mass term, 
\bqr  
F^{\a\b}{1\over\Box} F_{\a\b} = {1\over 4}\pa^{(\a}{}_{\dot\gamma} A^{\b)\dot\gamma} 
{1\over\Box_0} 
 \pa_{(\a\dot\rho} A^{\dot\rho}{}_{\b)} + \ldots = A^{\a\b} A_{\a\b} + \ldots \ , 
\fqr 
after integrating by parts.  The theory is a four-dimensional version of the chiral anomaly 
in two dimensions.  The quantum extension of \rf{massivetheory} has a different form that 
we next analyze.  Because the theory has a local form that is manifestly real, the theory 
is renormalizable and unitary (after projecting out the three negative norm states of 
the tensor fields $G^{\a\b}$ and $G^{\da\db}$).  The three-component anti-selfdual tensor 
$G_{\a\b}$ gives the degrees of freedom describing the massive vector.  One may consider 
processes with external $G_{\a\b}$ fields as required by unitarity; we integrate them out 
in the loop to generate the massive component of the gauge fields.  

The theory in four-vector notation has the form, 
\bqr  
{\cal L} ={\rm Tr}~ -{1\over 2} G^{\mu\nu} \Box G_{\mu\nu} - {1\over 4} F^{\mu\nu} 
F_{\mu\nu}  + M G^{\mu\nu} F_{\mu\nu} \ . 
\label{vectoraction} 
\fqr 
(We could write the gauge field kinetic term as $-{1\over 8} F^2 -{1\over 8} 
{\tilde F}^2$ as in 
\rf{localform}.)  We conclude this section with a discussion of the unitarity and the 
unphysical modes of the tensor field.  The action in \rf{vectoraction} has the linearized 
gauge invariance 
\bqr 
G^{\mu\nu} \rightarrow G^{\mu\nu} + \partial^\mu \theta^\nu - \partial^\nu \theta^\mu 
\fqr 
and 
\bqr  
A^\mu \rightarrow {1\over M} \Box \theta^\mu \ , 
\fqr 
upon restricting to 
\bqr  
M=-1 \qquad M^2\theta^\mu +{1\over 2} \Box \theta^\mu =0 \ .  
\fqr 
This invariance possesses three components as the latter equation restricts the vector 
to be massive.  The action is invariant under $G\rightarrow \lambda G$, $x^\mu 
\rightarrow \lambda^{-1} x^\mu$, $A^\mu\rightarrow \lambda A$, $M\rightarrow \lambda M$ 
and $g\rightarrow g$; as 
a result the invariance holds for all values of the mass parameter as this scaling 
results in $\lambda M=-1$.  The three components are sufficient to cancel the 
three modes of the anti-symmetric $G^{\mu\nu}$ tensor possessing negative norm, rendering 
the propagator well-defined.  After the dualization, one mode is absorbed by the gauge 
the field and two remain.  The gauge covariantization of this symmetry lifts to the 
full theory.  

\section{Quantum Extension} 

The quantum extension of \rf{massivetheory} is found by summing the one-particle reducible 
bubble diagrams and covariantizing.  It has the form, 
\bqr  
{\cal L} = {\rm Tr} ~{1\over 2} F^{\a\b} F_{\a\b} 
- {M^2\over 2} \sum_{k=0}^\infty  F^{\a\b} {1\over\Box} \Bigl( {M^2\over\Box} 
\ln{\Lambda^2/\Box}\Bigr)^k F_{\a\b} + {\rm h.c.}   
\nonumber 
\fqr 
\bqr \qquad\quad 
= {\rm Tr} ~{1\over 2} F^{\a\b} F_{\a\b} 
 - {M^2\over 2}~ F^{\a\b} {1\over\Box} \Bigl[ {1\over 1- {M^2\over\Box} 
\ln{\Lambda^2/\Box} }\Bigr] F_{\a\b} + {\rm h.c.} \ .
\label{covariantoneloop}
\fqr 
There are two regimes in which we may expand the denominator in \rf{covariantoneloop}.  
In the momentum range $k^2>>M^2$ we obtain the interaction from the series 
\bqr  
{\cal L}_{\rm int}^{(a)} = 
-{M^2\over 2}{\rm Tr} ~F^{\a\b} \sum_{n=0}^\infty {1\over \Box} \Bigl\{ {M^2\over\Box} 
 \ln{\Lambda^2/\Box}  \Bigr\}^n F_{\a\b} + {\rm h.c.} 
\label{smallenergy}
\fqr 
and in the range $M^2 >> k^2$, 
\bqr 
{\cal L}_{\rm int}^{(b)} = {\rm Tr}~ {1\over 2 \ln\left({\Lambda^2/\Box}\right)} ~ 
 F^{\a\b}\sum_{k=0}^\infty \Bigl\{ {\Box\over M^2} \ln{\Lambda^2\over\Box} \Bigr\}^k  
F_{\a\b} + {\rm h.c.} \ .  
\label{largeenergy} 
\fqr 
Both expanded forms permit a representation in diagrammatic rules.  The logarithmic 
terms in ${\cal L}_{\rm int}^{(a)}$ and ${\cal L}_{\rm int}^{(b)}$ are non-analytic 
and appear not to generate local interactions; however, they can be mapped into local 
interactions via analytic continuation or through the use of auxiliary fields.  At high 
energies, greater than $M^2$ the theory consists of massive vector bosons.  There is 
a Landau pole located to this order in the quantum expansion at approximately $k^2\sim -M^2 
\ln{(-M^2/k^2)}$.  

The expansion of the inverse of the covariant box is,  
\bqr  
{1\over\Box} &=& {1\over \Box_0 + \left\{ A^{\a\da},\pa_{\a\da} \right\} + 
A^{\a\da}A_{\a\da}  } 
\nonumber 
\fqr 
\bqr \qquad 
= {1\over\Box_0} \sum_{k=0}^\infty 
\left[ -{1\over\Box_0} \left( \left\{ A^{\a\da},\pa_{\a\da} \right\} + 
A^{\a\da}A_{\a\da}\right) \right]^k \ ,   
\fqr 
and generates the interactions in the second regime in which $k^2>>M^2$ from 
\rf{covariantoneloop} 
\bqr 
{\cal L}_1 = -{M^2\over 2}~ {\rm Tr} ~
 F^{\a\b} {1\over\Box_0} \sum_{k=1} \Bigl[ { \{ A^{\a\da},\pa_{\a\da} \} 
 + A^2 \over \Box_0 } \Bigr]^k  
\nonumber 
\fqr 
\bqr 
\times\sum_{p=0}^\infty \left[ {1\over \left[ 1-{M^2\over \Box_0} 
\ln{\Lambda^2/\Box_0}\right] }~   \sum_{k=0}^\infty M_k \right]^p F_{\a\b} 
\label{expansionone} 
\fqr 
with terms in $M_k$ containing products of $k$ gauge fields.  The complete theory 
includes also the interaction ${\rm Tr} ~F^{\a\b}F_{\a\b}+ {\rm Tr}~ F^{\da\db} 
F_{\da\db}$.  

We next expand the theory in a series in the gauge couplings in the first regime.  
The theory expanded to order ${\cal O}(A)$ contains the zeroeth order term, 
\bqr 
{\cal L} = {\rm Tr} ~F^{\a\b} F_{\a\b} 
- {M^2\over 2}~ F^{\a\b} {1\over \Box_0} \Bigl[ 
 {1\over 1-{\Lambda^2\over\Box_0} -\ln\Box_0/\Lambda^2}\Bigr] F_{\a\b} + {\rm h.c.} 
\fqr  
and the first order correction, 
\bqr  
{\cal L}_a = -{M^2\over 2} {\rm Tr} ~F^{\a\b} {1\over\Box_0} \Bigl({1\over 1- {M^2\over\Box_0} 
 \ln{\Lambda^2/\Box} }\Bigr) \bigl( -{1\over\Box_0} \{A,\pa\} \bigr) F_{\a\b} + {\rm h.c.}  \ . 
\fqr 
The expanded form to order ${\cal O}(A^2)$ has the form, 
\bqr \hskip -.2in
{\cal L}_b = {M^2\over 2} 
 {\rm Tr} ~	F^{\a\b} {1\over\Box_0} \Bigl[{1\over 1- {M^2\over\Box_0} 
\ln{\Lambda^2/\Box} }\Bigr] \nonumber 
\fqr 
\bqr \times
 \Bigl[ {M^2\over\Box_0^2} A^2 \ln{\Lambda^2\over\Box_0} - \bigl(\ln{\Lambda^2\over\Box_0}
\bigr) 
{M^2\over\Box_0^2}  \{A,\pa\} {1\over\Box_0} \{A,\pa\}  \Bigr] F_{\a\b} \ , 
\fqr 
together with the hermitian conjugate.  There are an infinite number of higher order 
terms in the expansion but these vertices are sufficient to generate the four-point function.  

\section{Diagrammatic Rules} 

We first specify the line factors associated with the massive vector bosons 
\cite{spinortwo,Kife}.  There are three independent polarizations that satisfy the 
completeness relation, 
\bqr  
\sum_{\lambda=\pm,0} \epsilon_{\a\da}^\star(k,M) \epsilon_{\b\db}(k,M) = C_{\a\b} C_{\da\db} 
 + {k_{\a\da} k_{\b\db}\over 2 M^2} \ .  
\fqr 
The polarizations written in terms of spinor inner products of legs with an arbitary massive 
momentum $k_{\a\da} = k_{-,\a\da}+k_{+,\a\da}$, and $k_+^2=k_-^2=0$ (so that $k_\pm^{\a\da} 
= k_\pm^\a k_\pm^\da$), have the explicit form, 
\bqr 
\epsilon_{+}^{\a\da} (k;M) = {k_+^\a k_-^\da \over M} \qquad\qquad \epsilon_-^{\a\da} (k;M) = 
 {k_-^\a k_+^\da \over M} \ , 
\fqr  
and 
\bqr  
\epsilon_0^{\a\da}(k;M) = {1\over \sqrt{2} M} \left( k_+^\a k_+^\da - k_-^\a k_-^\da \right) 
\ , 
\fqr 
and momentum in spinor form,  
\bqr 
M^2=-k^2 = \langle k_+ k_-\rangle [k_- k_+] \ .  
\fqr 
These leg factors have the ambiguity in the representation of breaking the off-shell momentum 
into two null vectors, which may be utilized to simplify calculations.  

The local form of the massive theory \rf{deformedgauge} has the Feynman rules, with 
the gauge-fixing term for the $A^{\a\da}$ field specified with ${\cal L}_{\rm gf}= 
\lambda^2/2 (\pa\cdot A)^2$ and $\lambda=1$, consisting of the propagators, 
\bqr  
\langle G^{\a\b}(k) G^{\mu\nu}(-k)\rangle = {1\over k^2+m_0^2} \left( C^{\a\mu} C^{\b\nu}+ 
C^{\a\nu}C^{\b\mu}  \right) \ ,
\nonumber 
\fqr 
\bqr 
\langle A^{\a\db}(k) A^{\b\dot\gamma}(-k) \rangle = {1\over k^2} C^{\a\b} 
C^{\db\dot\gamma} \ ,  
\fqr 
and the two-point interaction, 
\bqr  
\langle G^{\a\b}(k) A^{\mu\dot\nu}(-k) \rangle = M k^{(\a|\dot\nu} C^{\b)}{}_\mu \ .
\nonumber 
\fqr 
Summing the diagrams with the $\langle AG\rangle$ vertex generates the mass 
term for $A$.  The three- and four-point vertices are, 
\bqr 
\langle A^{\mu\dot\nu}(k_1) A^{\a\db}(k_2) G^{\gamma\delta}(k_3) \rangle = 
 C^{\dot\nu\dot\beta} \left( C^{\mu\nu} C^{\alpha\delta} + C^{\mu\delta} 
C^{\alpha\gamma} \right) 
\nonumber 
\fqr 
\bqr 
\langle A^{\a_1\da_1}(k_1) A^{\a_2\da_2}(k_2) A^{\a_3\da_3}(k_3) \rangle = 
 k_3^{\dot\beta_3(\alpha_1} C^{\a_2\a_3} C^{\db_1\db_2} + {\rm perms} \ ,
\fqr 
and
\bqr 
\langle \prod_{j=1}^4 A^{\a_j\da_j}(k_j) \rangle = C^{\db_1\db_2} C^{\db_3\db_4} 
 \left( C^{\a_1\a_3} C^{\a_2\a_4} + C^{\a_2\a_3} C^{\a_1\a_4} \right) + {\rm perms} \ . 
\fqr 
These interactions generate the local form of the dualized massive theory.  

The quantum one-loop effective action following from integrating out the auxiliary $G^{\a\b}$ 
fields in the two-point function and covariantizing is listed in \rf{covariantoneloop}.  These 
vertices follow from a resummation of one loop diagrams.  The propagator is,  
\bqr  \hskip -.5in 
V_2 = \Bigl( -{1\over 2} \left\{ 1+ 11{M^2\over k^2+ M^2 \ln(-k^2/\Lambda^2)} \right\} 
 \left( k^{\a\db} k^{\b\da} -{1\over 4} k^2 C^{\a\b} C^{\da\db} \right)   
\fqr 
\bqr 
+ {1\over 2} \left\{ 1+9 {M^2\over k^2 + M^2\ln(-k^2/\Lambda^2)} \right\} \left( 
 k^{\a\db} k^{\b\da} +{k^2\over 4} C^{\a\b} C^{\da\db} \right) \Bigr)^{-1}
\fqr 
The third variation of the Lagrangian generates the color ordered three-point vertices, 
\bqr  \hskip -.6in
V_3^{(a)} = \lambda\left[ k_1^{\da_1\a_2} C^{\da_2\da_3} C^{\a_1\a_3} + 
k_2^{\da_2\a_3} C^{\da_3\da_1} C^{\a_2\a_1} 
 + k_2^{\da_3\a_1} C^{\da_1\da_2} C^{\a_3\a_2} \right] 
\label{threepointa} 
\fqr 
\bqr \hskip -.6in
V_3^{(b)} = {M^2\over k_1^2+ M^2\ln{\left(-\Lambda^2/k_1^2\right)}} ~
 {1\over k_1^2} \left[ k_1^{\da_1(\a} C^{\b)\a_1} k_3^{\da_3}{}_{\a} 
 C_{\b}{}^{\a_4} {1\over k_1^2} \left( k_3+2k_2\right)^{\a_2\da_2} \right] 
\label{threepointb} 
\fqr 
\bqr \hskip -.2in
V_3^{(c)} = -{1\over 2} {M^2\over k_1^2+ M^2\ln{\left(-\Lambda^2/k_1^2\right)} }~ \Bigl[ 
 k_1^{\da_1\a_2} C^{\da_2\da_3} C^{\a_1\a_3} + k_2^{\da_2\a_3} C^{\da_3\da_1} C^{\a_2\a_1} 
\nonumber 
\fqr 
\bqr
 + k_2^{\da_3\a_1} C^{\da_1\da_2} C^{\a_3\a_2} \Bigr] 
\label{threepointc}
\fqr 
and the fourth, the color ordered four-point vertices, 
\bqr  
\hskip -.3in 
V_4^{(a)} = \Bigl[ 
 {\lambda^2\over 2} -{1\over 2} {M^2\over (k_2+k_3)^2 + M^2 
\ln\left( -{\Lambda^2/k_1^2}\right)} \Bigr] 
 C^{(\a_1|\a_2} C^{\a_3)a_4} C^{(\da_1|\da_2} C^{\da_3)\da_4} 
\label{fourpointa} 
\fqr 
\bqr \hskip -.2in 
V_4^{(b)} = \Bigl[ {M^2\over k_1^2 + M^2 \ln\left(-{\Lambda^2/k_1^2}\right)} \Bigr] \Bigl\{
\left({1\over k_1^2}- {M^2\over k_1^4}\right) k_1^{\da_1(\a} C^{\b)\a_1} k_4^{\da_4}{}_{(\a} 
 C_{\b}{}^\a_4 C^{\a_2\a_3} C^{\da_2\da_3} 
\nonumber 
\fqr 
\bqr 
 - \bigl(\ln{\Lambda^2\over k_1^2}\bigr) {M^4\over k_1^4} {1\over (k_3+k_4)^2} 
\left( k_2+2k_3+2k_4\right)^{\a_2\da_2} \left( k_4+2k_3\right)^{\a_3\da_3} \Bigr\} 
\label{fourpointc} 
\fqr 
The ordering of the inverse boxes follows from the expansion of the one-loop 
effective action.  There are an infinite number of higher-point interaction vertices 
in the fully expanded form; for simplicity we derive the vertices up to the four-point 
order.  

\section{Summary}  

In this work we examine a selfdual inspired reformulation of a theory of 
massive vector bosons found by perturbing a Lorentz covariant selfdual theory 
\cite{Chalmers:1996rq} to  
\bqr  
{\cal L} = {\rm Tr} ~{1\over 2} G^{\a\b} \Box G_{\a\b} + M G^{\a\b} F_{\a\b} + 
{1\over 2} G^{\da\db} \Box G_{\da\db} + MG^{\da\db} F_{\da\db} 
\nonumber 
\fqr 
\bqr 
+ {1\over 2} {\rm Tr} ~\left( F^{\a\b} F_{\a\b} + F^{\da\db} F_{\da\db}\right) \ , 
\label{localrep}
\fqr 
in which the selfdual theory is described by ${\cal L}={\rm Tr} ~G^{\a\b} F_{\a\b}$.  
At low-energies the dynamics is governed by an exactly solvable selfdual system and 
at higher energies, above the energy scale set by the dimensionful coupling $M$, the 
theory consists of massive vector bosons in which the additional degrees of freedom 
are absorbed by the auxiliary field imposing the selfduality constraint.  The 
Lagrangian \rf{localrep} possesses a gauge symmetry.  This symmetry is sufficient 
to gauge away half of the components of the tensor fields and allows for unitary 
propagation of the modes. 

In previous works selfdual inspired reformulations of gauge theories has led to 
improved Feynman rules and expansions \cite{Chalmers:1996rq}.  These include second 
order reformulations of fermionic couplings \cite{Morgan:1995te} 
and spontaneously broken gauge theories \cite{Chalmers:2001cy}.  
In this work the additional degrees of freedom required to compose a massive vector 
are replaced by a Lagrange multiplier field that imposes the selfduality constraint 
in the undeformed case.  The Lagrangian we consider in this work has a non-local 
form with an expansion containing an infinite number of interactions; it is an analog 
of the chiral anomaly in two dimensions.  Furthermore the theory permits a local form 
\rf{localrep} in which the renormalizibility and unitarity properties are manifest.  
We examine the interactions up to four-point order.  

\vskip .3in 
\section*{Acknowledgements} 

The work of GC is supported in part by the US Department of Energy, Division of High 
Energy Physics, contract W-31-109-ENG-38.  GC thanks Cosmas Zachos and the referee 
for helpful comments.

\end{document}